\documentclass[12pt, preprint]{aastex}
\usepackage[dvips]{color}

\def\ltsima{$\;\buildrel < \over \sim \;$}
\def\simlt{\lower.5ex \hbox{\ltsima}}
\def\gtsima{$\;\buildrel > \over \sim \;$}
\def\simgt{\lower.5ex \hbox{\gtsima}}

\newcommand{\re}{\textcolor[rgb]{0,0,0}}

\shorttitle{Collisional rate coefficients from an artificial neural network}
\shortauthors{Neufeld}

\begin{document}

\title{Rate coefficients for the collisional excitation of molecules: \\
estimates from an artificial neural network}
\author{David A.~Neufeld}
\affil{Department of Physics and Astronomy, Johns Hopkins University,
3400~North~Charles~Street, Baltimore, MD 21218}

\begin{abstract}

An artificial neural network (ANN) is investigated as a tool for estimating rate coefficients for the collisional excitation of molecules.  The performance of such a tool can be evaluated by testing it on a dataset of collisionally-induced transitions for which rate coefficients are already known: the network is trained on a subset of that dataset and tested on the remainder.  Results obtained by this method are typically accurate to within a factor $\sim 2.1$ (median value) for transitions with low excitation rates and $\sim 1.7$ for those with medium or high excitation rates, although 4$\%$ of the ANN outputs are discrepant by a factor of 10 more.  The results suggest that ANNs will be valuable in extrapolating a dataset of collisional rate coefficients to include high-lying transitions that have not yet been calculated.  For the asymmetric top molecules considered in this paper, the favored architecture is a cascade-correlation network that creates 16 hidden neurons during the course of training, with 3 input neurons to characterize the nature of the transition and one output neuron to provide the logarithm of the rate coefficient.

\end{abstract}


\section{Introduction}

The interpretation of astrophysical spectra rests heavily upon the availability of relevant atomic and molecular data.  Of critical importance are (1) the energies of the various accessible quantum states, (2) the rates of spontaneous radiative decay between states, and (3) the state-to-state rate coefficients for excitation and deexcitation in inelastic collisions.  The last of these have proven quite elusive; in many circumstances, our ability to interpret astrophysical spectra is severely limited by the availability of collisional rate coefficients.  As a result of new spectroscopic facilities such as the {\it Spitzer Space Telescope}, the {\it Herschel Space Observatory}, and the planned Atacama Large Millimetric Array (ALMA), the need for fundamental molecular data has become all the more acute.  In modeling the emission from interstellar molecular clouds, in particular, there is a critical need for rate coefficients for the rotational excitation of molecules in collisions with H$_2$.  The theoretical calculation of such data is extremely time consuming, involving the computation of a potential energy surface for the system in question, followed by extensive scattering calculations.  Even for those molecules for which such calculations have been carried out, the available data are often limited to collisionally-excited collisions involving a relatively small set of low-lying states.  Submillimeter and infrared spectra of molecular clouds often reveal the presence of higher-lying transitions for which collisional rate coefficients have not been computed; here, some kind of extrapolation method is needed.  For the case of linear molecules, the infinite order sudden (IOS) and related approximations provide an obvious method of extrapolation; for asymmetric molecules, there is no simple analogous method.

Given a limited set of transitions of an asymmetric molecule for which collisional rate coefficients have been computed, previous efforts at extrapolation (e.g.\ Neufeld \& Melnick 1987, Faure \& Josselin 2008) have typically adopted an expression with one or more adjustable parameters for how the collisional rate coefficient depends upon the relevant parameters for each transition (energy difference, change in rotational quantum number, etc.)  The adjustable parameters are then chosen so as to optimize the fit to those transitions for which collisional rate coefficients are known, and then the expression is used to estimate the values for higher-lying transitions where calculated results are unavailable.  For example, the recent study by Faure \& Josselin (2008) considered the rate coefficients for the excitation of H$_2$O by H$_2$, which will be of crucial importance to the interpretation of many {\it Spitzer} and {\it Herschel} spectra.  Here, Faure \& Josselin extrapolated a set of rate coefficients obtained from quasi-classical trajectory calculations for transitions among the lowest 45 rotational states of water. Each such state is characterized by an energy, $E$, and rotational quantum numbers $J$, $K_a$ and $K_c$ that describe the total angular momentum and its projection onto the principal axes of the molecule.
For transitions with $\Delta K_a \le 2$, $\Delta K_c \le 2$ and $\Delta J \le 2$ -- termed ``high-propensity" transitions -- Faure \& Josselin assumed that rate coefficients depend solely upon $\Delta K_a$, $\Delta K_c$ and $\Delta J$.  For all other transitions (``low propensity transitions"), they assumed the rate coefficients to be {\it independent} of $\Delta K_a$, $\Delta K_c$ and $\Delta J$ and proportional to $\exp (-\Theta \Delta E/kT)$, where $\Theta$ is adjusted to optimize the fit to the available data.

As an alternative to the methods adopted previously, I have explored the use of an artificial neural network (ANN) to accomplish the extrapolation.  In contrast to previous techniques, the use of an ANN does not make any {\it a priori} assumption about how the transitions might be categorized or how the rate coefficients might depend upon the parameters for each transition. The calculational method is discussed in \S2 (with the more technical details given in the Appendix).  The performance of the ANN in reproducing a known set of collisional rate coefficients in presented in \S3.  A discussion follows in \S4.

\section{Calculational method}

Since the mid-1980's, artificial neural networks have been widely used to solve a wide variety of problems, particularly in pattern recognition.   Previous astrophysical applications have centered primarily on the automatic classification of astronomical objects based on morphology, photometry or spectroscopy \re{(e.g.\ Ball et al.\ 2004, Collister \& Lahav 2004), although ANN have also been used previously as approximation tools (e.g. Asensio Ramos \& Socas-Navarro 2005).}  

ANNs consist of a set of interconnected artificial neurons.  Such neurons are of three types: input neurons, output neurons, and hidden neurons.   
Given a set of input values that are fed into the input neurons, a unique set of output values emerges from the output neurons. In the context of the application considered here, the input neurons are fed the relevant parameters for each transition (e.g. the energies and relevant quantum numbers for the upper and lower states), and the collisional rate coefficient emerges from a single output neuron.  The hidden neurons mediate the dependence of the ANN's output upon its input.

Each individual neuron accepts a particular input value, and outputs a value that is -- in general -- a non-linear function of that input.  This function is called the activation function, and is typically sigmoidal in shape.  (The inverse hyperbolic tangent is an often-used activation function in ANN.)
With the exception of the input neurons, for which the inputs are a set of input parameters,  the input for any neuron is the weighted sum of the outputs of a set of other neurons to which it is connected.  During the ``learning process", the ANN is shown a "training set" of input parameters, together with the appropriate outputs; the relevant weights of the connections are varied until the outputs (from the output neurons) match the appropriate values to within the desired accuracy.  

The arrangement of the connections between the neurons is called the ``architecture" of the network.  In ``feed-forward" networks such as the one used in this study, the structure is non-recursive: this means that, when considered in the appropriate order, the input value for any neuron reflects only output values for neurons that have already been considered.  In some ANN, the architecture is fixed, and the learning process simply involves varying the weights of the connections.  In other schemes, such as the cascade-correlation network introduced by Fahlman and Lebiere (1990) and used in the present study, the architecture evolves during the learning process, with hidden neurons trained and added one at a time to improve the performance of the ANN.

Figure 1 shows a simple example of a cascade-correlation network, with 3 input neurons (blue circles numbered 1 -- 3), 2 hidden neurons (numbered 4 and 5) , and one output neuron (number 6).  The $i$th neuron has an input value $x_i$ (shown in red) and an output value $y_i$ = $f(x_i)$ (green), where $f$ is an activation function (which may, in principle, be different for each neuron -- see Appendix).    Information propagates from left to right, starting with the input values $x_1$, $x_2$ and $x_3$, and ending with the output value $y_6$.   The $x_i$ and $y_i$ are related by a matrix of weights, {\bf W}, with $x_i = \Sigma\, W_{ij} y_j$.  In the example shown here, {\bf W} has 12 non-zero elements.   At the end of training, the weights have assumed values that optimize the performance on the training set: in other words, $y_6$ is close to the desired output value for each given set of inputs, $[x_1,x_2,x_3]$.

In this study, I used the Fast Artificial Neural Network (FANN) Library, an open source software package developed by Steffen Nissen and Evan Nemerson\footnote{available at http://fann.sourceforge.net}.  Full details of my use of this software are presented in the Appendix; my intent there is to provide enough specificity to allow the reader to reproduce the calculational method exactly.  A copy of the relevant computer code, written in C, will be provided upon request to the author.

In evaluating the performance of an ANN, it is essential to train the network on a ``training set" of data and then test its performance on a ``test set" {\it that it has never seen.}  Given sufficient complexity in the network (i.e. enough hidden neurons), it is always possible to fit the training set to any desired degree of accuracy.  The key question is how well the ANN performs on new, unseen data.  In this study, I considered the collisional rate coefficients computed recently by Dubernet et al.\ 
(2009)\footnote{available on the BASECOL database (Dubernet et al.\ 2006; http://basecol.obspm.fr)} for the excitation of ortho-H$_2$O in collisions with para-H$_2$.  The close-coupling calculations that led to these results represent the state-of-the-art in scattering calculations and are extremely expensive computationally.  This dataset provides the rate coefficients for all transitions among the 45 lowest rotational states of ortho-H$_2$O, computed for several temperatures and several values of the initial and final rotational states of the colliding para-H$_2$ molecule.  For simplicity, I considered first the values obtained at a single temperature, 800~K, and with the initial and final H$_2$ states being $J=0$.  This dataset then contains $\onehalf N_{tot} (N_{tot} - 1) = 990$ independent values, where $N_{tot}=45$ is the total number of water rotational states.  The factor of $\onehalf$ arises because the upwards and downwards rates for any transition are related by considerations of detailed balance and are therefore not independent of each other.  As the training set, I adopted a subset of these 990 data points; \re{because the ultimate goal of this method is to estimate collisional rate coefficients for transitions that are {\it higher} in energy than those calculated previously, I chose a training set consisting of those collisional-induced transitions between the {\it lowest} $N_{train}$ rotational states,} where $N_{train} < N_{tot}$.  The training set then consists of  $\onehalf N_{train} (N_{train} - 1)$ data points, with the test set comprising the remaining  $\onehalf N_{tot} (N_{tot} - 1) - \onehalf N_{train} (N_{train} - 1)$ data points.

One key methodological choice concerns the input parameters that will be provided to the ANN; this determines the number of input neurons.  Each rotational state of ortho-H$_2$O is characterized by an energy $E$, and three rotational quantum numbers: $J$, which describes the total rotational angular momentum, $K_a$, which describes its projection on the principal axis for which the moment of inertia is smallest, and $K_c$, which describes its projection on the principal axis for which the moment of inertia is largest.  However, once $J$ and $K_a$ are specified, $K_c$ is uniquely determined.  Thus it is convenient to define a parameter $\tau = K_a- K_c$, which allows each rotational state of ortho-H$_2$O to be specified unambiguously by just two quantum numbers, $J$ and $\tau$.  The integer $\tau$ takes all odd values between $-J$ and $+J$.  In the present study, I experimented with three choices for the input parameter set.  The three cases considered were: (1) the ANN is fed just the energy difference between the initial and final state, $\Delta E$, and therefore has one input neuron; (2) the ANN is fed $\Delta E$,  $\Delta J$, and $\Delta \tau$, and therefore has 3 input neurons; (3) the ANN is fed $E$, $J$, and $\tau$ for both the initial and final states, requiring a total of 6 input neurons.  In \S3 below, the performance of the ANN is described for each of these three cases.  In all cases, the ANN has single output neuron which provides the rate coefficient for the downward collisionally-induced transition.  Because the 
rate coefficients cover several orders of magnitude, I trained the network to output the logarithm of the rate coefficient.

Two other choices, more technical in nature, determine the performance of the ANN.  The final number of hidden neurons, $N_{hidden}$, is discussed in the Appendix.  Based upon the experimentation described there, I adopted $N_{hidden}=16$ as the optimum value.  With too few hidden neurons, the ANN fails to fit even the training data set well; with too many hidden neurons, the network fits the training set perfectly but fails to generalize well to the test set.  Because the initial weights on the connections are randomly assigned, the results vary slightly every time the ANN is trained.  Accordingly, I took the average of the outputs of several ANN in obtaining results for the test set.  As discussed in Appendix A, the performance of the ANN fails to improve once the number of realizations, $N_r$, exceeds $\sim 10$; \re{this behavior indicates that the remaining errors in the ANN predictions are not a consequence of the particular choice of initial weights, but are instead a fundamental limitation of the method.} The results presented in \S 3 below all apply to an ANN with $N_{hidden}=16$ and $N_r = 10$

\section{Performance}

\subsection{Dependence upon the input parameters provided to the ANN}

The top right, bottom left, and bottom right panels in Figure 2 present the performance of an ANN with $N_{train} = 30$ for the cases with 1, 3, and 6 input neurons discussed in \S2 above.   In each panel, the horizontal axis represents the logarithm of the rate coefficient (in units of $\rm cm^3\,s^{-1}$) obtained by Dubernet et al.\ (2009), and the vertical axis represents the corresponding output of the ANN.  Red points apply to the 435 data points in the training set, while blue points apply to the 555 data points in the test set (i.e.\ to those transitions that were not seen by the ANN during training).
As expected from the discussion of high and low propensity transitions presented by Faure \& Josselin (2008), the ANN performs much better for the cases with 3 and 6 input neurons; only in these cases, is it informed about the rotational quantum numbers along with the transition energy.  Indeed, the performance in the 1-input-neuron is strikingly similar to the results obtained if the data are all fit with an expression of the form $k \exp(-\Theta\Delta E/kT)$.  Adopting best-fit values of $k$ and $\Theta$ determined from a least squares fit to the data yields an expression which is compared to the actual data in the top left panel of Figure 2.

More detailed statistical information is provided in Figures 3 and 4.  In Figure 3, the frequency distribution of the errors, $\rm log_{10} (ANN\,\, output/actual\,\,value)$, is shown for each 
case.\footnote{The present study addresses only the ability of the ANN to reproduce the Dubernet et al.\ (2009) results (and not the accuracy of those results); thus the Dubernet et al.\ (2009) results will be referred to here as the ``actual" rate coefficients.}   Black histograms apply to the entire test set, while blue, green and red histograms apply to transitions with low, medium and high actual rate coefficients; for this purpose, ``low" applies to value smaller than $10^{-13} \rm \, cm^3\, s^{-1}$, ``medium" to value in the range $10^{-13} -- 10^{-11} \rm \, cm^3\, s^{-1}$, and ``high" to a value greater than $10^{-11} \rm \, cm^3\, s^{-1}$.  Clearly, the cases with 3 and 6 neurons yield histograms that are more strongly peaked than that obtained with 1 input neuron, indicating that the typical error in the ANN output is considerably smaller.  While the probability distributions for the 3- and 6-input-neuron cases are very similar when all the test data are considered together (black histogram), the 6-input-neuron ANN systematically underpredicts the results for transitions with high rate coefficients (red histogram; this behavior is also apparent from a careful inspection of Figure 2.)  Paradoxically, the ANN performs worse when it is provided with additional information (i.e. the individual $E$, $J$, $\tau$ values for the upper and lower states rather than just the differences between those values, $\Delta E$, $\Delta J$, $\Delta \tau$).  In the 6-input-neuron case, the ANN is 
\re{evidently fitting a trend with energy, $E$, that is present for the lower energy transitions (i.e.\ for the training set), but extrapolates poorly to the higher-energy transitions encountered in the test set.} 

Based upon Figures 2 and 3, I adopted the 3-input-neuron ANN as the optimum architecture, along with the values $N_{hidden}=16$ and $N_r=10$ discussed in the Appendix.  For this architecture, the distribution of the output errors is presented in an alternative form in Figure 4. Here I show the fraction of the data for which $\rm \vert log_{10}(ANN\,\,output/actual\,\,value) \vert $ exceeds a given value, $x$.  The fraction of rate coefficients is shown on the vertical axis (logarithmic scale), with the same color coding adopted in Figure 3, as a function of $x$.  For example, the fact that the red curve passes through the point (0.19 ,--0.30) implies that one-half (i.e.\ 10$^{-0.3}$) of those rate coefficients with high actual values (i.e. greater than $10^{-11} \rm \, cm^3\, s^{-1}$) are in error by more than a factor of 1.5 (i.e.\ $10^{0.19}$) Once again, results apply only to the test set.  Clearly, the ANN output is less reliable for transitions with rate coefficients smaller than $10^{-13} \rm \, cm^3\, s^{-1}$ (blue curve).  Even in this case, however, more than half of the rate coefficients are correct to within a factor 2.1

The results shown in Figures 2 -- 4 were all obtained with $N_{train}=30$, for which the training set comprises the lowest 435 transitions and the test set contains the next 555 transitions.  I have also investigated the performance of the ANN as a function of the size of the training set, i.e.\ as a function of $N_{train}$.  In Figure 5, I show the RMS value of $\rm log_{10}(ANN\,\,output/actual\,\,value)$, as a function of N$_{train}$, for the entire set of transitions (black points) and for low, medium and high values of the actual rate (blue, dark green and red as before).  Results are shown for $N_{train} = 10$, 15, 20, 25, 30, 35, and 40.  These correspond to training sets containing 45, 105, 190, 300, 435, 595, and 780 data points, respectively, and test sets containing 945, 785, 800, 690, 555, 395, and 210 data points.  The overall performance varies little once $N_{train} \ge 20$, i.e.\ once the training set encompasses $\simgt 20 \%$ of the total available data.   

One key application of the present study, of course, will be to train the ANN on all 990 transitions for which Dubernet et al.\ have computed collisional rate coefficients, and to then use the ANN to estimate the rates for higher-lying transitions for which no calculations have been performed.  
The results of Figure 5 provide a gauge of the likely success of that procedure, and of the number of additional rate coefficients that can be estimated reliably.  If the accuracy of the extrapolation depends upon the {\it relative} size of the training set \re{(i.e.\ the {\it ratio} of the number of transitions in the training set to the total number for which ANN estimates are to be obtained)}, we may expect that an additional $\sim 4000$ rate coefficients can be predicted with the use of an ANN.  On the other hand, it is possible that the {\it absolute} size of the training set is the relevant parameter; in that case, there is no obvious limit to how far the extrapolation can be carried.  A conservative approach would be to use an ANN to extrapolate no more than an additional $\sim 4000$ rate coefficients; the resultant data set would include all transitions among the lowest $\sim 100$ rotational states of o-H$_2$O.  For higher-lying transitions, the method of Faure \& Josselin (2008) promises to be more robust: in that method, the results for high-propensity transitions are obviously well-behaved, and those for low-propensity  transitions tend gracefully to zero as $\Delta E$ tends to infinity.  With the ANN, by contrast, the behavior could be unstable if one extrapolates to a $\Delta E$ (or $\Delta J$ or $\Delta \tau$) that is much larger than anything it saw in the training set.

\subsection{Other cases}

While the collisional excitation of water has received a great deal of theoretical attention recently, many other molecules will be observed by ALMA and Herschel.  Figure 6 shows results obtained for three other molecules for which there is a large set of collisional rate coefficients in the BASECOL database: SO$_2$ (from Green 1995), ortho-c-C$_3$H$_2$ (Chandra \& Kegel 2000), and ortho-H$_2$CO (Green 1991).  Like water, these are asymmetric top molecules with rotational states that can be characterized by the quantum numbers $J$ and $\tau$, although in these cases, the available data apply to excitation in collisions with He.  I have also obtained additional results for the collisional excitation of ortho-water with para-H$_2$, again using rate coefficients from Dubernet et al.\ (2009) but now for cases in which the initial and/or final state of H$_2$ is $J=2$.   In each case, the training set consisted of the lower-lying half of the entire available data set.  The RMS value of $\rm log_{10}(ANN\,\,output/actual\,\,value)$ for the test set is printed on each panel; as in the case of ortho-H$_2$O colliding with H$_2$ ($J=0\rightarrow 0$), typical RMS values are roughly 0.35 

\section{Summary}

1.  Artificial neural networks (ANN) can be successfully trained to determine rate coefficients for the collisional excitation of astrophysical molecules.

2.  The type of ANN favored here is a cascade correlation network which creates 16 hidden neurons during the course of training.  There are three input neurons that inform the network of $\Delta E$, $\Delta J$ and $\Delta \tau$ for a given transition, and one output neuron that provides the logarithm of the rate coefficient for collisional de-excitation.

3.  The performance of such an ANN can be evaluated by training it on a subset of the set of transitions for which collisional rate coefficients are known, and then testing it on the remainder of that set.  Since the motivation for this study is to develop a method for estimating the rate coefficients for high-lying transitions for which values have not been computed, the training set involved the lowest-lying transitions among the available data.

4.  The performance of the ANN increases slightly if the results from $\sim 10$ realizations are averaged.

5.  In the primary test case considered here, the excitation of ortho-H$_2$ by H$_2$ ($J=0$), the available data -- consisting of 990 rate coefficients -- can be modeled equally well provided that the network is trained on at least $20\%$ of the available data. 

6.  The performance of the ANN is summarized in Table 1, which lists the typical output errors in every case considered.  Here the RMS and median values of $\rm log_{10}(ANN\,\,output/actual$ value) are listed for the entire test set and separately for low, medium and high values of the actual rate.  Overall, for ANN with 16 hidden neurons and 3 input neurons that are trained on at least 20$\%$ of the available data, the typical RMS value of $\rm log_{10}(ANN\,\,output/actual\,\,value)$ is $\sim 0.4$; for transitions with rate coefficients $\le 10^{-13} \rm cm^{-3}\,s^{-1}$, the typical RMS value is $\sim 0.5$.  The corresponding median values of $\vert \rm log_{10}(ANN\,\,output/actual\,\,value) \vert$ are $\sim 0.25$ and $\sim 0.33$.  These median values imply that one-half of the ANN outputs are typically accurate to within factors of 1.8 (or $\sim 2.1$ in the case of rate coefficients $\le 10^{-13} \rm cm^{-3}\,s^{-1}$.)   However, the results shown in Figure 4 indicate that some of the ANN outputs are highly discrepant (with 4$\%$ in error by more than a factor 10). It should be noted that the errors given here apply to individual state-to-state rate coefficients.  When these are used to model molecular line emissions, the errors in the resultant line strengths can be expected to be smaller than in the state-to-state rate coefficients because a given radiative transitions may be pumped by a combination of several different collisionally-induced transitions. 

\vskip 0.5 true in

\centerline{\bf Appendix: Calculational details}

The calculation made use of the Fast Artificial Neural Network (FANN) Library, version 2.0.0, an open source software package developed by Steffen Nissen and Evan Nemerson and available at http://fann.sourceforge.net.  The cascade training option was adopted for this problem, and was found to result in rapid training.  In this algorithm (Fahlman \& Lebiere 1990), the training starts with an empty network devoid of hidden neurons.   Hidden neurons are trained and added to the network one at a time; each new hidden neuron is chosen from a selection of candidate neurons with different initial weights and one of several possible activation functions. 

In using the FANN library, the input neuron values were rescaled to \re{small values (relative to unity) to ensure that the input neurons did not saturate}: energies (given as wavenumber in units of cm$^{-1}$) were therefore divided by $10^5$ and rotational quantum numbers were divided by 100.  \re{No other preprocessing was performed.  With these rescalings, the input values lay in the ranges [0, 0.14], [--0.03, 0.10], and [--0.16, 0.18] respectively for $\Delta E$, $\Delta J$, and $\Delta \tau$.}
The desired output neuron values were the logarithm \re{to base 10} of the rate coefficient in units of $\rm cm^{3}\,s^{-1}$, \re{and covered the range [--15.3, --9.9]} 

The cascade algorithm was used with all the default settings except for one: a linear function was selected as the activation function for the output layer.  This choice was necessitated by the fact that the output neuron values lay outside the range of the other activation functions\footnote{\re{An alternative solution would have been to rescale the output neuron values.}}.  The desired error was set to a very small value to ensure that the ANN made use of the maximum number of hidden neurons specified for each test.
\re{The default settings for the FANN software make six activation functions available to the algorithm, all of which are continuous functions with a domain of [$-\infty,+\infty$] and a range of either [0, 1] or [--1, 1].  The network is optimized using the i-RPROP algorithm of Igel \& H\"usken (2000), a variant of the RPROP algorithm introduced by Riedmiller \& Braun (1993).}

Even if the settings and training set are unchanged, the results are slightly different every time the ANN is run.  This is because the initial weights are randomly assigned at the start of training.  I therefore averaged the output values over a set of $N_r$ runs.  Figure 7 shows how the performance of the network depends upon $N_r$, for a network with 3 input neurons and 16 hidden neurons and with $N_{train}=30$.  As in Figures 3 -- 5, black curves apply to the entire test set, while blue, green and red histograms apply to transitions with low, medium and high actual rate coefficients; for this purpose, ``low" applies to value smaller than $10^{-13} \rm \, cm^3\, s^{-1}$, ``medium" to value in the range $10^{-13} - 10^{-11} \rm \, cm^3\, s^{-1}$, and ``high" to a value greater than $10^{-11} \rm \, cm^3\, s^{-1}$. The magenta curve refers to the training set.  Clearly, the performance on the test set is improved if multiple runs are averaged, but for $N_r$ greater than a few, no further improvements are realized.  Based upon the behavior exhibited in Figure 7, I selected $N_r = 10$ as the standard parameter for the calculations that yielded the results discussed in \S3. 

Figures 8 and 9 show how the performance depends upon the final number of hidden neurons.  The color coding in Figure 9 is the same as that in Figure 7.   Clearly, if enough hidden neurons are provided, the fit to the training set becomes extremely accurate (lower right panel in Figure 8), with the RMS error for 64 neurons being less than 3$\%$.  However, the performance on the {\it test} set worsens somewhat once the number of hidden neurons exceeds 16.  This behavior indicates that the network is ``overfitting" the data (e.g.\ Smith 1993).  The following analogy is presented by the use of polynomial fitting functions.  Given a function of one variable that has been evaluated at $N$ values, a perfect polynomial fit to those values can always be obtained with a polynomial of order $N-1$.  However, such a polynomial may show strong wiggles between (and beyond) the points where the function was evaluated (particularly if the function evaluations have errors).  A lower order polynomial may fit the data points imperfectly but may be preferable for the purposes of interpolation and particularly extrapolation. 
Based upon the behavior exhibited in Figure 8 and 9, I selected $N_{hidden} = 16$ as the standard parameter for the calculations that yielded the results discussed in \S3.

\acknowledgments

I thank George Fekete for helpful discussions about ANN, and Alexandre Faure for several valuable comments on an earlier version of the manuscript.  \re{I am grateful to the anonymous referee for several useful suggestions.}
I gratefully acknowledge the support of funding from Research Support Agreement 1349625 issued by NASA's Jet Propulsion Laboratory.

\begin{deluxetable}{lllllllllllllll}
\rotate
\tablewidth{0pt}
\tablecaption{Summary of ANN performance} 
\tablehead{$N_{hidden}$ & $N_{input}$ & $N_{train}$ & $N_{test}$ & Molecule & Collider & $T$/K & \multicolumn{4}{c}{RMS $\rm log_{10}(output/actual)$} &
\multicolumn{4}{c}{Median $\vert \rm log_{10}(output/actual) \vert$}}

\startdata
     &      &      &      &              &             &      & all & low$^a$ & med.\ $^a$ & high$^a$ &  all & low$^a$ & med.\ $^a$ & high$^a$ \\
   2 &    3 &  435 &  555 &        o-H$_2$O &  H$_2$  0--0$^b$ &  800 & 0.483 & 0.652 & 0.394 & 0.425 & 0.284 & 0.374 & 0.259 & 0.292 \\
   4 &    3 &  435 &  555 &        o-H$_2$O &  H$_2$  0--0 &  800 & 0.461 & 0.634 & 0.373 & 0.359 & 0.266 & 0.386 & 0.237 & 0.208 \\
   8 &    3 &  435 &  555 &        o-H$_2$O &  H$_2$  0--0 &  800 & 0.417 & 0.527 & 0.366 & 0.352 & 0.239 & 0.279 & 0.227 & 0.178 \\
  16 &    3 &  435 &  555 &        o-H$_2$O &  H$_2$  0--0 &  800 & 0.417 & 0.517 & 0.371 & 0.350 & 0.247 & 0.327 & 0.227 & 0.186 \\
  32 &    3 &  435 &  555 &        o-H$_2$O &  H$_2$  0--0 &  800 & 0.446 & 0.559 & 0.391 & 0.411 & 0.259 & 0.345 & 0.235 & 0.207 \\
  64 &    3 &  435 &  555 &        o-H$_2$O &  H$_2$  0--0 &  800 & 0.471 & 0.569 & 0.422 & 0.473 & 0.288 & 0.364 & 0.235 & 0.176 \\
\\
  16 &    3 &   45 &  945 &        o-H$_2$O &  H$_2$  0--0 &  800 & 0.915 & 1.322 & 0.809 & 0.538 & 0.458 & 0.784 & 0.433 & 0.286 \\
  16 &    3 &  105 &  885 &        o-H$_2$O &  H$_2$  0--0 &  800 & 0.675 & 1.007 & 0.561 & 0.491 & 0.358 & 0.473 & 0.314 & 0.423 \\
  16 &    3 &  190 &  800 &        o-H$_2$O &  H$_2$  0--0 &  800 & 0.474 & 0.653 & 0.416 & 0.299 & 0.291 & 0.402 & 0.279 & 0.143 \\
  16 &    3 &  300 &  690 &        o-H$_2$O &  H$_2$  0--0 &  800 & 0.409 & 0.531 & 0.364 & 0.305 & 0.237 & 0.306 & 0.225 & 0.152 \\
  16 &    3 &  435 &  555 &        o-H$_2$O &  H$_2$  0--0 &  800 & 0.417 & 0.517 & 0.371 & 0.350 & 0.247 & 0.327 & 0.227 & 0.186 \\
  16 &    3 &  595 &  395 &        o-H$_2$O &  H$_2$  0--0 &  800 & 0.391 & 0.430 & 0.377 & 0.307 & 0.227 & 0.264 & 0.222 & 0.139 \\
  16 &    3 &  780 &  210 &        o-H$_2$O &  H$_2$  0--0 &  800 & 0.381 & 0.447 & 0.341 & 0.328 & 0.214 & 0.258 & 0.199 & 0.123 \\
\\
  16 &    3 &  435 &  555 &        o-H$_2$O &  H$_2$  0--0 &  800 & 0.417 & 0.517 & 0.371 & 0.350 & 0.247 & 0.327 & 0.227 & 0.186 \\
  16 &    6 &  435 &  555 &        o-H$_2$O &  H$_2$  0--0 &  800 & 0.406 & 0.441 & 0.353 & 0.691 & 0.293 & 0.352 & 0.249 & 0.615 \\
  16 &    1 &  435 &  555 &        o-H$_2$O &  H$_2$  0--0 &  800 & 0.725 & 0.906 & 0.613 & 0.889 & 0.420 & 0.466 & 0.393 & 0.829 \\
\\
\\
\\
\\
     &      &      &      &              &             &      & all & low$^a$ & med.\ $^a$ & high$^a$ &  all & low$^a$ & med.\ $^a$ & high$^a$ \\
  16 &    3 &  495 &  495 &        o-H$_2$O &  H$_2$  0--0$^b$ &  800 & 0.433 & 0.577 & 0.353 & 0.372 & 0.234 & 0.314 & 0.209 & 0.239 \\
  16 &    3 &  495 &  495 &        o-H$_2$O &  H$_2$  0--2 &  800 & 0.338 & 0.675 & 0.276 & 0.361 & 0.195 & 0.450 & 0.174 & 0.224 \\
  16 &    3 &  495 &  495 &        o-H$_2$O &  H$_2$  2--0 &  800 & 0.325 & 0.368 & 0.274 & N/A & 0.188 & 0.229 & 0.163 & N/A \\
  16 &    3 &  495 &  495 &        o-H$_2$O &  H$_2$  2--2 &  800 & 0.342 & 0.457 & 0.329 & 0.368 & 0.210 & 0.319 & 0.202 & 0.221 \\
\\
  16 &    3 &  540 &  541 &  o-c-C$_3$H$_2$ &     He       &   30 & 0.325 & N/A & 0.269 & 0.414 & 0.208 & N/A & 0.171 & 0.316 \\
  16 &    3 &  602 &  603 &          SO$_2$ &     He       &   25 & 0.341 & 0.361 & 0.327 & 0.369 & 0.205 & 0.220 & 0.187 & 0.368 \\
  16 &    3 &  346 &  346 &       o-H$_2$CO &     He       &   30 & 0.288 & 0.636 & 0.270 & 0.221 & 0.168 & 0.646 & 0.155 & 0.163 \\

\enddata

\tablenotetext{a}{Statistics for the subset of test data with actual rate coefficients that are low ($\le 10^{-13} 
\,\rm cm^3\,s^{-1}$),  medium ($10^{-11} - 10^{-13}  \,\rm cm^3\,s^{-1}$) or high ($\ge 10^{-11}  \,\rm cm^3\,s^{-1}$) }

\tablenotetext{b}{The notation H$_2$ X--Y indicates that the initial H$_2$ state is $J=$X and the final state is $J=$Y}

\end{deluxetable}

\begin{figure}
\includegraphics[scale=0.70]{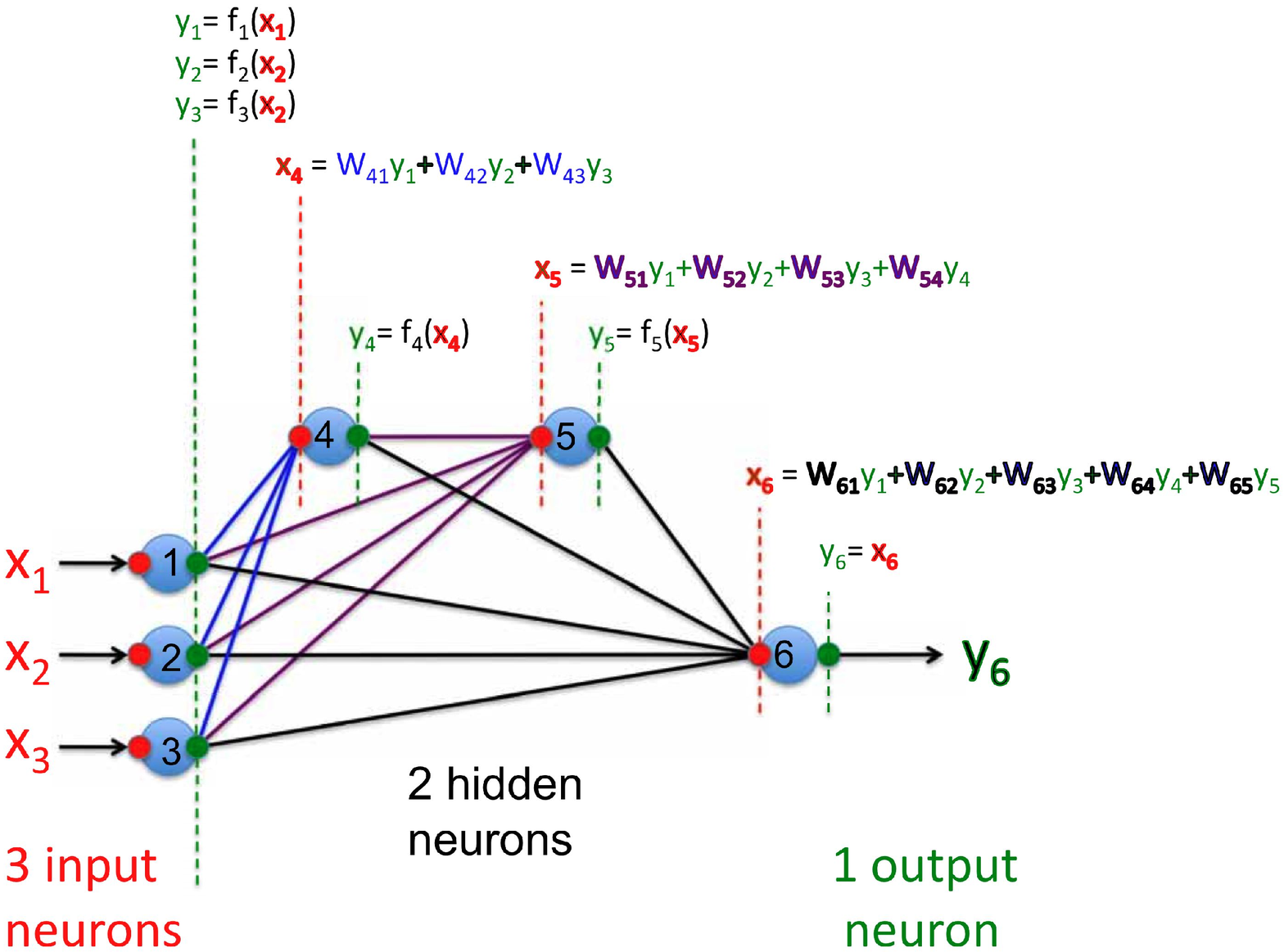}

\noindent{Fig.\ 1 -- Simple example of a cascade-correlation network, with 3 input neurons (blue circles numbered 1 -- 3), 2 hidden neurons (numbered 4 and 5) , and one output neuron (number 6).  The $i$th neuron has an input value $x_i$ (red) and an output value $y_i$ = $f(x_i)$ (green), where $f$ is an activation function (which may, in principle, be different for each neuron -- see Appendix).    Information propagates from left to right, starting with the input values $x_1$, $x_2$ and $x_3$, and ending with the output value $y_6$.   The $x_i$ and $y_i$ are related by a matrix of weights, {\bf W}, with $x_i = \Sigma\, W_{ij} y_j$.  The algebraic expressions shown on the figure show that the quantities $y_1$, $y_2$, $y_3$, $x_4$, $y_4$, $x_5$, $y_5$, $x_6$, and $y_6$ can be computed non-recursively (i.e.\ explicitly, in the above order), given input values of $x_1$, $x_2$, and $x_3$.}
\end{figure}

\begin{figure}
\includegraphics[scale=0.80]{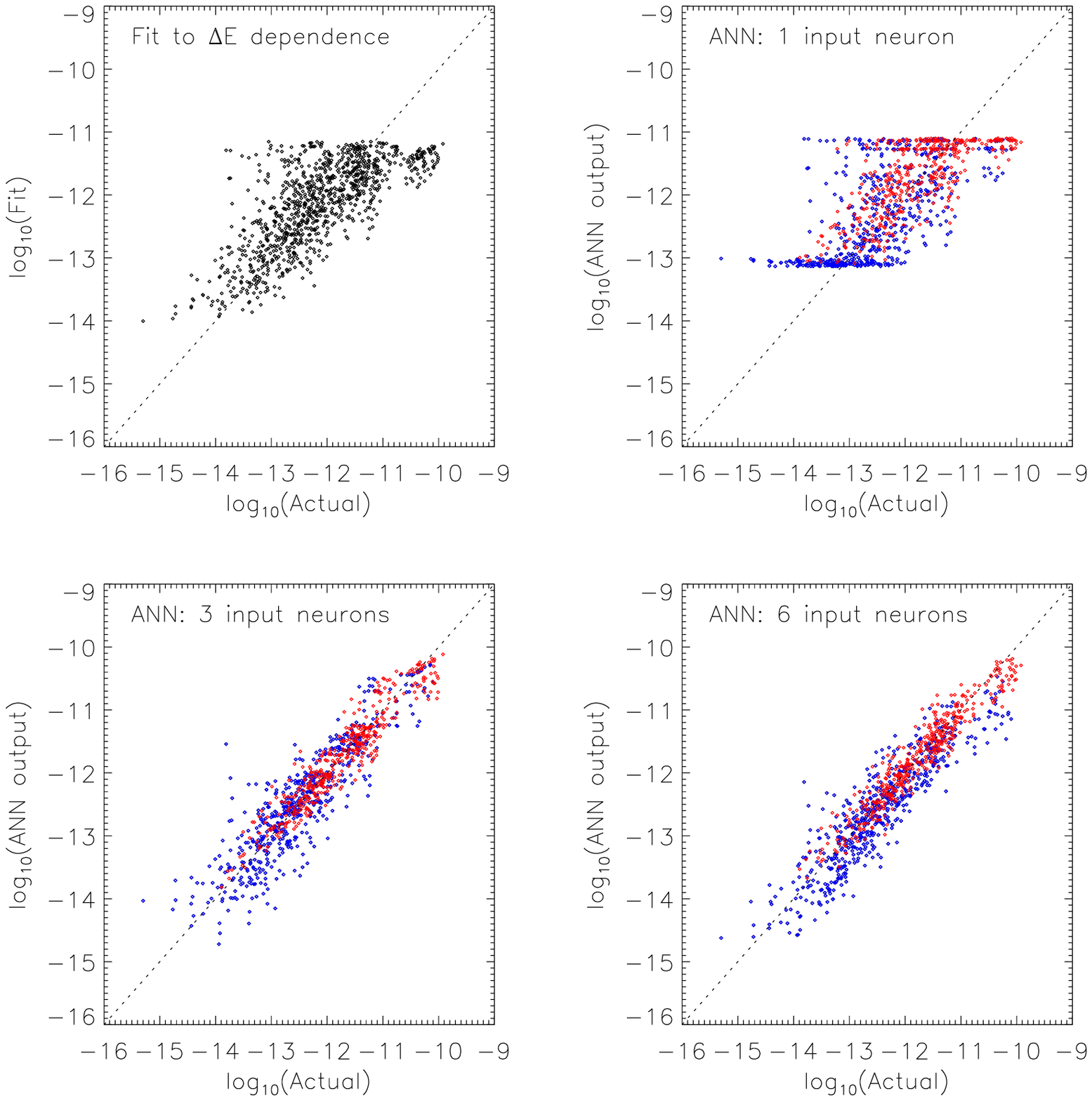}

\noindent{Fig.\ 2 -- The top right, bottom left, and bottom right panels present the performance of an ANN with $N_{train} = 30$ for the cases with 1, 3, and 6 input neurons discussed in \S2.   In each panel, the horizontal axis represents the logarithm of the rate coefficient (in units of $\rm cm^3\,s^{-1}$) obtained by Dubernet et al.\ (2009), and the vertical axis represents the corresponding output of the ANN.  Red points apply to the 435 data points in the training set, while blue points apply to the 555 data points in the test set (i.e.\ to those transitions that were not seen by the ANN during training).  Upper right panel: analogous results if the data are all fit with an expression of the form $k \exp(-\Theta\Delta E/kT)$.}
\end{figure}

\begin{figure}
\includegraphics[scale=0.95]{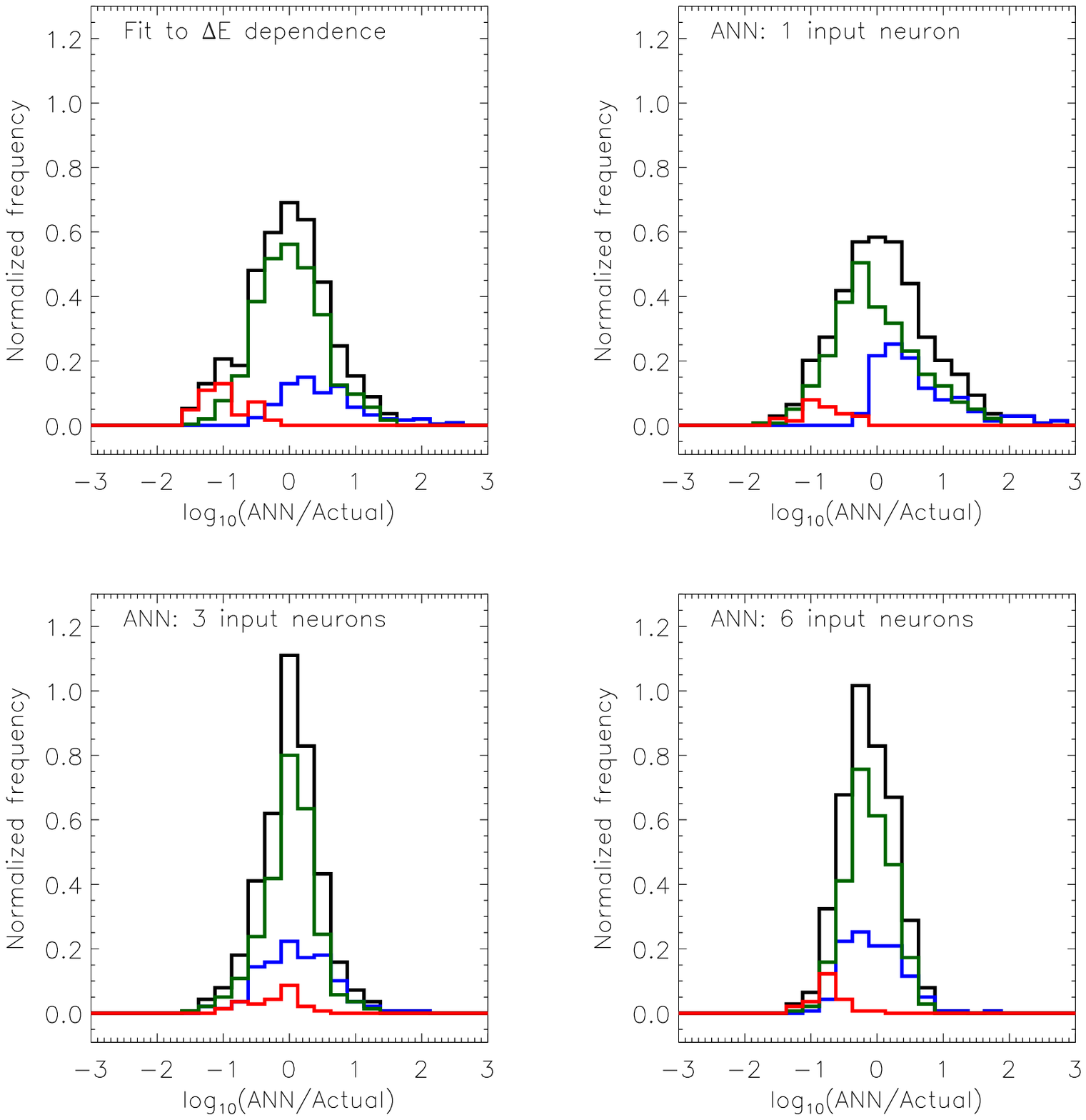}

\noindent{Fig.\ 3 -- The frequency distribution of the errors, $\rm log_{10} (ANN\,\, output/actual\,\,value)$, for each of the 4 cases shown in Fig.\ 1.  Black histograms apply to the entire test set, while blue, green and red histograms apply to transitions with low, medium and high actual rate coefficients; for this purpose, ``low" applies to value smaller than $10^{-13} \rm \, cm^3\, s^{-1}$, ``medium" to value in the range $10^{-13} - 10^{-11} \rm \, cm^3\, s^{-1}$, and ``high" to a value greater than $10^{-11} \rm \, cm^3\, s^{-1}$.}
\end{figure}

\begin{figure}
\includegraphics[scale=0.85]{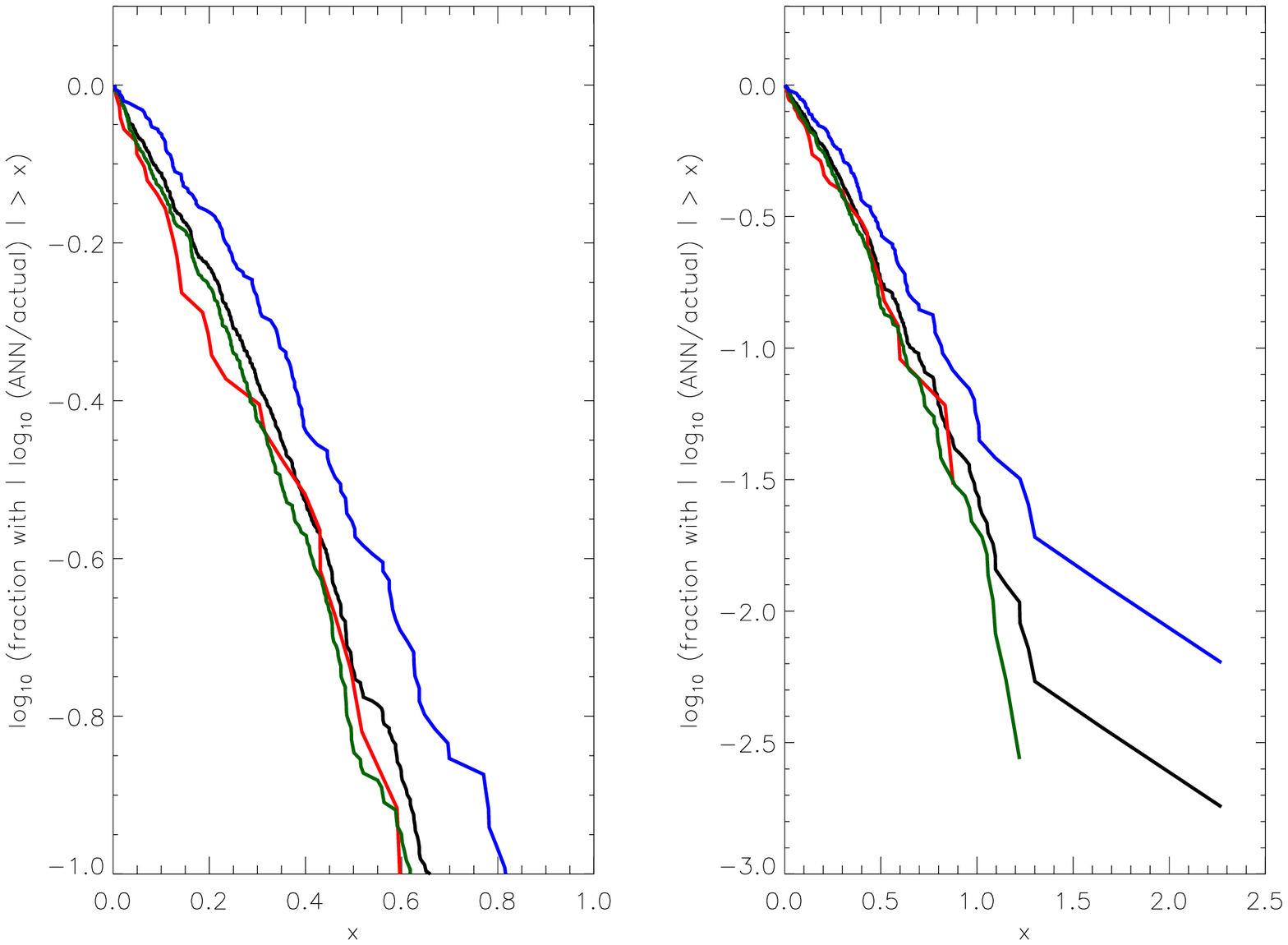}

\noindent{Fig.\ 4 --  Fraction of the test data for which $\rm \vert log_{10}(ANN\,\,output/actual\,\,value) \vert $ exceeds a given value, $x$.  The fraction of rate coefficients is shown on the vertical axis (logarithmic scale), with the same color coding adopted in Figure 2, as a function of $x$.  The left panel is simply a zoomed-in version of the right panel. }
\end{figure}

\begin{figure}
\includegraphics[scale=0.95]{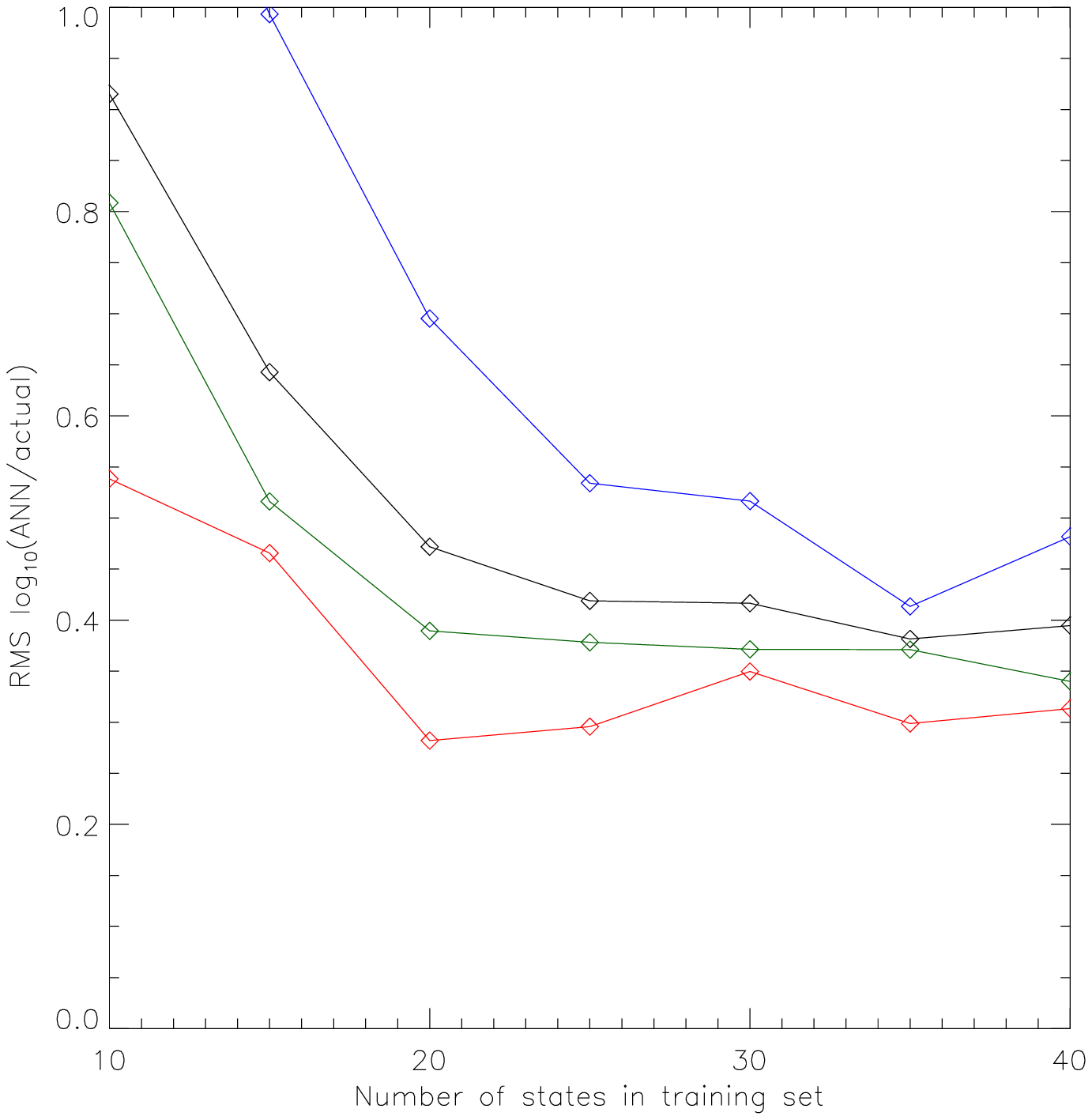}

\noindent{Fig.\ 5 -- RMS value of $\rm log_{10}(ANN\,\,output/actual\,\,value)$, as a function of N$_{train}$, for the entire set of transitions (black points) and for low, medium and high values of the actual rate (blue, dark green and red as in Figs.~2 and 3).}
\end{figure}

\begin{figure}
\includegraphics[scale=0.80]{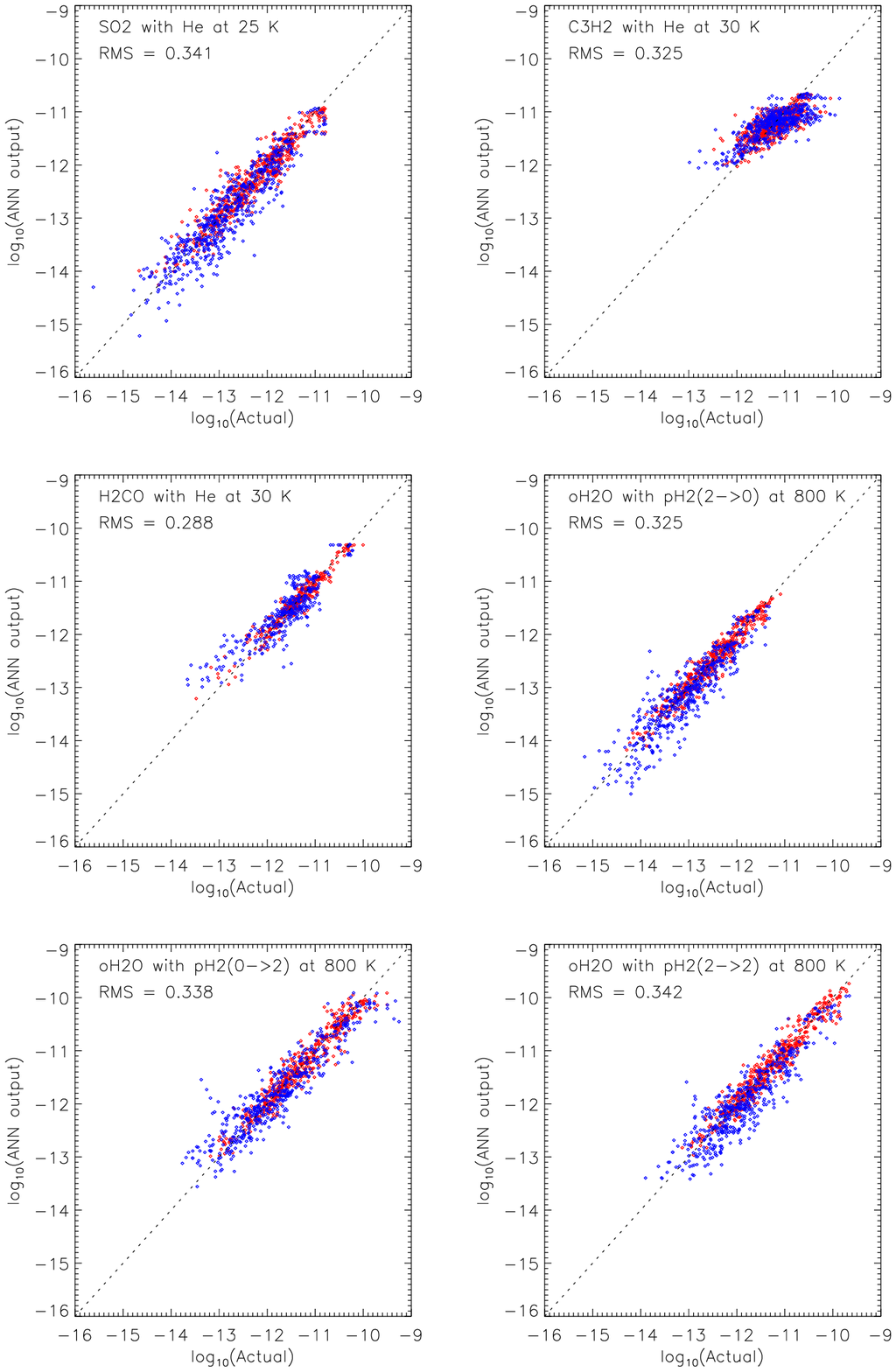}

\noindent{Fig.\ 6 -- Results obtained for excitation of SO$_2$ (from Green 1995), ortho-c-C$_3$H$_2$ (Chandra \& Kegel 2000), and ortho-H$_2$CO (Green 1991) in collisions with He; and for excitation of ortho-water in collisions with para-H$_2$ (Dubernet et al.\ (2009) for cases in which the initial and/or final state of H$_2$ is $J=2$.  }
\end{figure}

\begin{figure}
\includegraphics[scale=0.80]{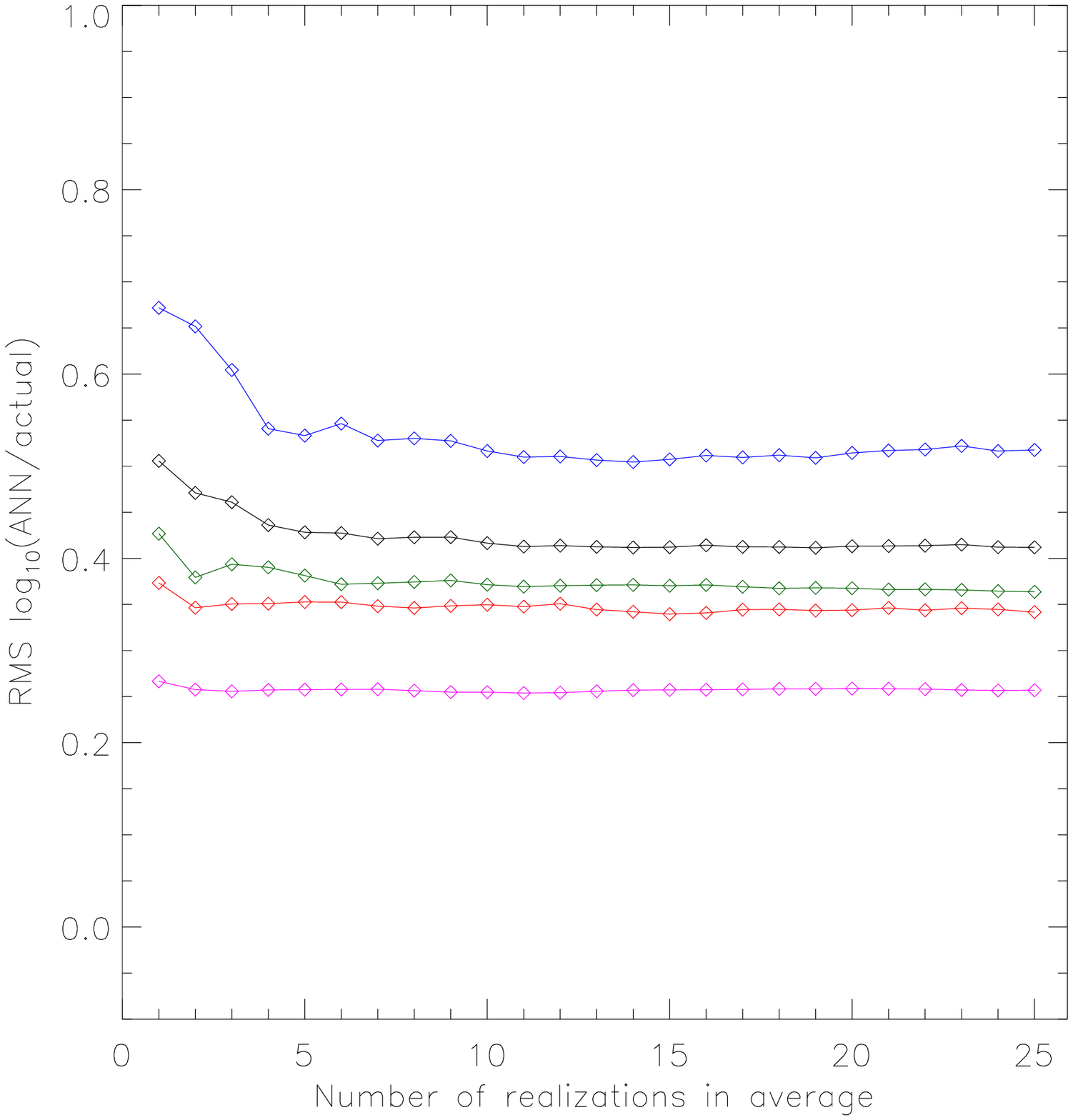}

\noindent{Fig.\ 7 -- Dependence of the performance of the ANN upon $N_r$, for a network with 3 input neurons and 16 hidden neurons and with $N_{train}=30$. \re{Black curves apply to the entire test set, while blue, green and red histograms apply to transitions with low, medium and high actual rate coefficients; for this purpose, ``low" applies to value smaller than $10^{-13} \rm \, cm^3\, s^{-1}$, ``medium" to value in the range $10^{-13} - 10^{-11} \rm \, cm^3\, s^{-1}$, and ``high" to a value greater than $10^{-11} \rm \, cm^3\, s^{-1}$. The magenta curve refers to the training set.}}
\end{figure}

\begin{figure}
\includegraphics[scale=0.80]{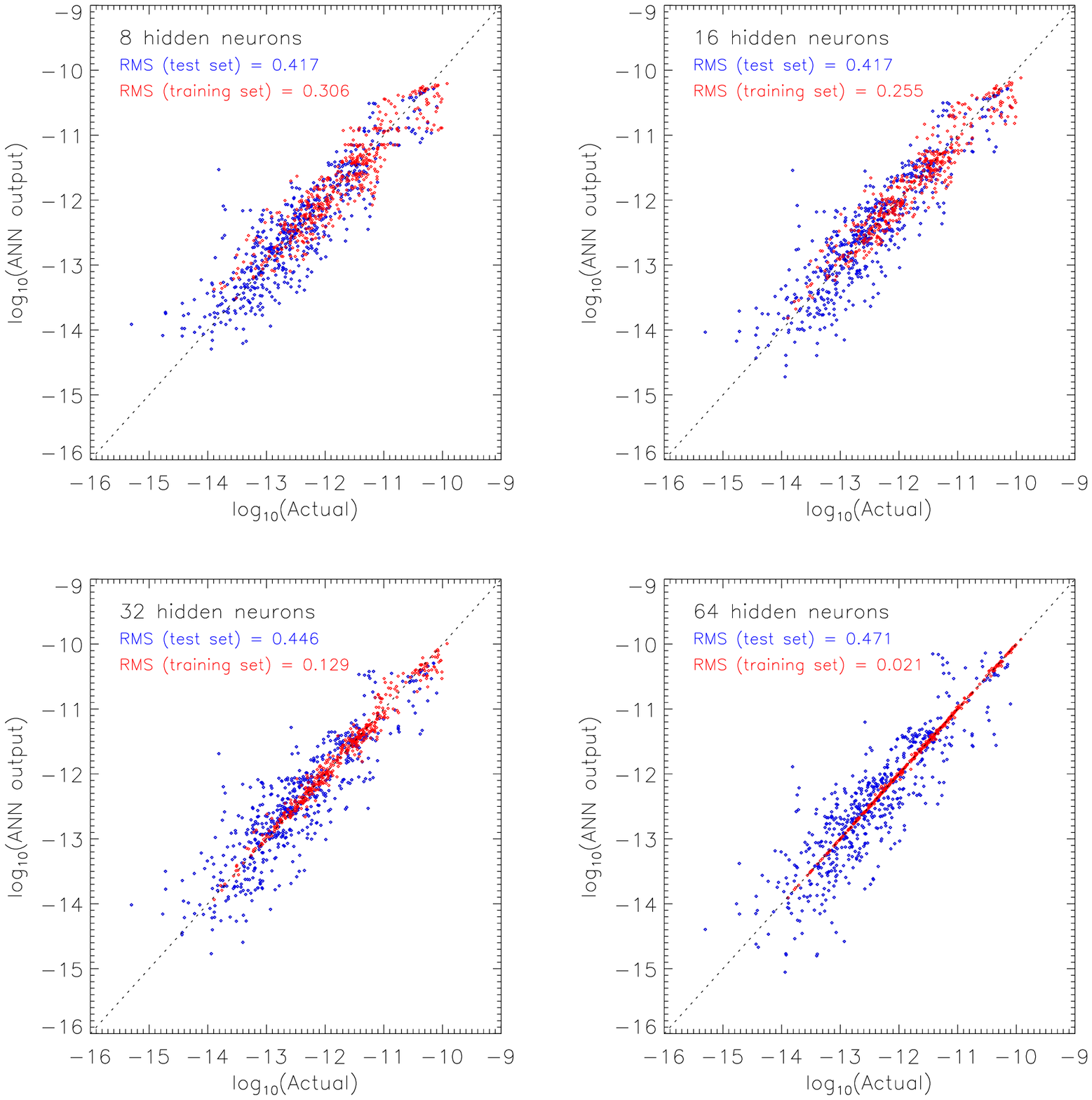}

\noindent{Fig.\ 8 -- Performance of an ANN with 3 input neurons and $N_{train}=30$.  Four panels shows the results for 8, 16, 32, and 64 hidden neurons.  In each panel, the horizontal axis represents the logarithm of the rate coefficient (in units of $\rm cm^3\,s^{-1}$) obtained by Dubernet et al.\ (2009), and the vertical axis represents the corresponding output of the ANN.  Red points apply to the 435 data points in the training set, while blue points apply to the 555 data points in the test set (i.e.\ to those transitions that were not seen by the ANN during training). }
\end{figure}

\begin{figure}
\includegraphics[scale=0.80]{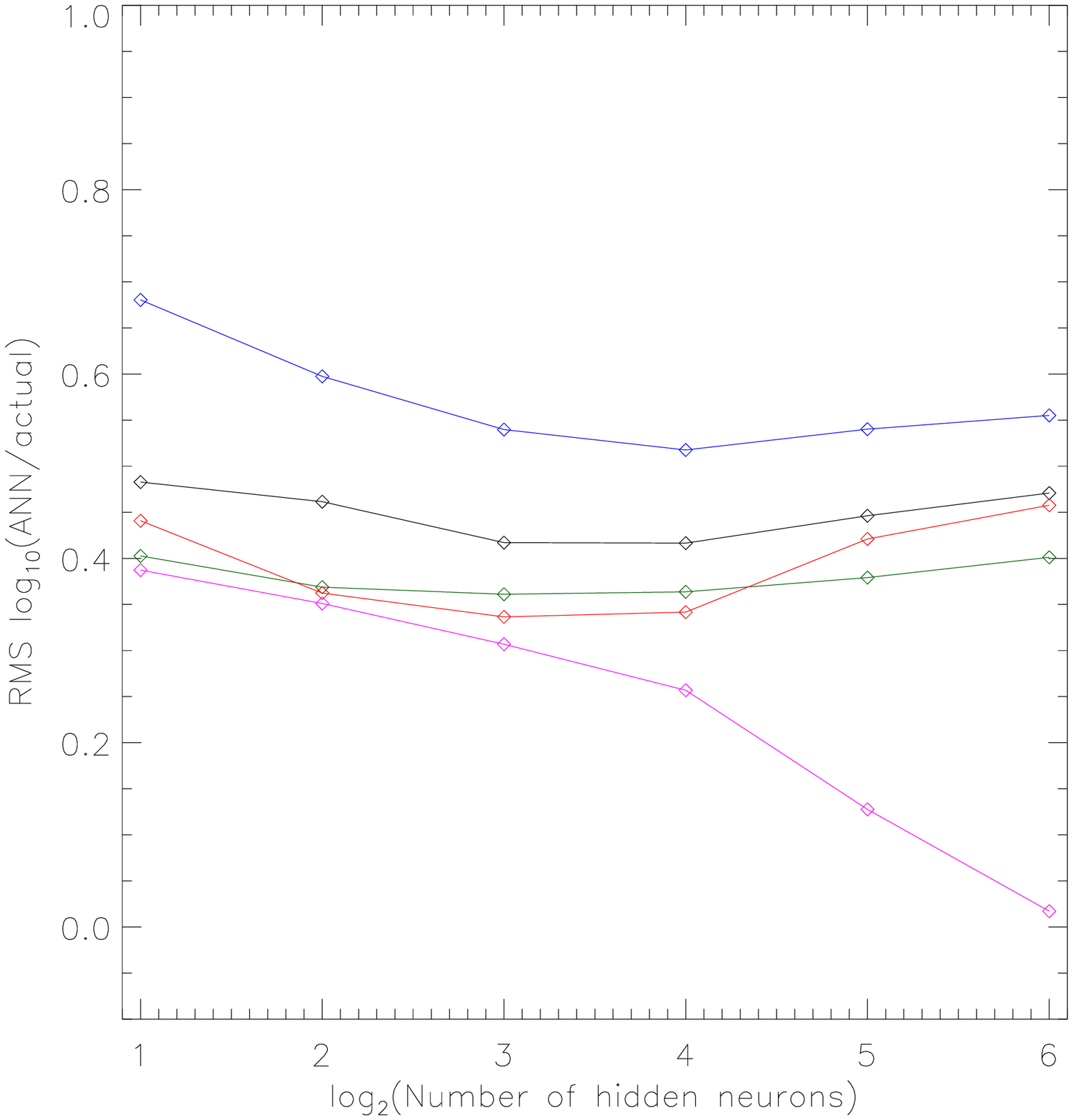}

\noindent{Fig.\ 9 -- Dependence of the performance of the ANN upon the number of hidden neurons, for a network with 3 input neurons and $N_{train}=30$.  Black curves apply to the entire test set, while blue, green and red histograms apply to transitions with low, medium and high actual rate coefficients; for this purpose, ``low" applies to value smaller than $10^{-13} \rm \, cm^3\, s^{-1}$, ``medium" to value in the range $10^{-13} - 10^{-11} \rm \, cm^3\, s^{-1}$, and ``high" to a value greater than $10^{-11} \rm \, cm^3\, s^{-1}$. The magenta curve refers to the training set.  }
\end{figure}

\end{document}